\documentclass[12pt]{article}
\usepackage{epsfig}

\usepackage{url}
\usepackage{slashed}
\usepackage{bm}
\usepackage{amsfonts}

\newcommand{\bo}{\partial^2}

\def\beq{\begin{equation}}
\def\eeq#1{\label{#1}\end{equation}}
\def\eeqn{\end{equation}}

\def\beqa{\begin{eqnarray}}
\def\eeqa#1{\label{#1}\end{eqnarray}}
\def\eeqan{\end{eqnarray}}

\let\bar=\overbar

\def\etal{{\it et al.}}

\def\eg{{\it e.g.}}

\def\Dslash{\not{\hbox{\kern-4pt $D$}}}
\def\dslash{\not{\hbox{\kern-2pt $\del$}}}

\def\msb{{\bar{\ssstyle M \kern -1pt S}}}

\def\Title#1{\begin{center} {\Large {\bf #1} } \end{center}}

\begin{document}

\Title{Low-energy limit of QCD at finite temperature}

\bigskip\bigskip

%+\addtocontents{toc}{{\it D. Reggiano}}
%+\label{ReggianoStart}

\begin{raggedright}  

{\it Marco Frasca\index{Frasca, M.}\\
Via Erasmo Gattamelata, 3\\
I-00176 Roma, ITALY}
\bigskip\bigskip
\end{raggedright}

\section{Introduction}

Lattice computations have provided evidence for the existence of a phase transition in QCD \cite{Fodor:2001pe,Aoki:2006we,Aoki:2009sc,Bazavov:2009zn} notwithstanding the sign problem casts some doubt about \cite{deForcrand:2010ys}. From a theoretical standpoint, a missing low-energy limit obtained from QCD forced people to use some relevant models to cope with this question as the Nambu-Jona-Lasinio model or the sigma model. Starting from these, some authors proposed the introduction of an imaginary chemical potential \cite{Alford:1998sd,de Forcrand:2002ci,D'Elia:2002gd}. Statistical confinement was proved in this framework and account was given for the phase diagram of QCD \cite{Fukushima:2008wg, Abuki:2008nm}. Recently, a different approach to identify a critical endpoint with a chiral chemical potential was also proposed \cite{Ruggieri:2011xc}.

The aim of this contribution is to show how, provided a recent analytical approach to manage low-energy QCD, it is possible to give a proof of existence of a critical point for chiral symmetry in a limit at zero quark masses and chemical potential. The idea is to fix the form factor in a non-local Nambu-Jona-Lasinio model as discussed in \cite{Hell:2008cc} by obtaining it directly from QCD and deriving in this way the proper low-energy limit of the theory. Finally, we will be able to obtain a critical temperature in close agreement with lattice data.

Our strategy will be to start from an analysis of classical field theory in the limit of a strong coupling and then, with the knowledge acquired in this way, we will be able to formulate a quantum field theory in this same limit. Extending the analysis to finite temperature will be accomplished with standard techniques. We will closely follow Ref.\cite{Frasca:2011bd}.

\section{Classical field theory}

A classical field theory for a massless scalar field is given by
\begin{equation}
   \bo\phi+\lambda\phi^3=j.
\end{equation}
The homogeneous equation can be solved exactly by
\begin{equation}
   \phi = \mu\left(\frac{2}{\lambda}\right)^\frac{1}{4}{\rm sn}(p\cdot x+\theta,i)
\end{equation}
being {\rm sn} an elliptic Jacobi function and $\mu$ and $\theta$ two constant. This solution holds provided the following dispersion relation holds
\begin{equation} 
   p^2=\mu^2\sqrt{\frac{\lambda}{2}}
\end{equation}
so this solution represents a free massive solution notwithstanding we started from a massless theory. Mass arises from the nonlinearities when $\lambda$ is taken to be finite rather than going to zero as normally done in weak perturbation theory. In this latter case, it would be very difficult to recover the above solutions. When there is a current we ask for a solution in the limit $\lambda\rightarrow\infty$ as our aim is to understand a strong coupling limit. So, we check a solution
\begin{equation}
   \phi=\kappa\int d^4x'G(x-x')j(x')+\delta\phi
\end{equation}
being $\delta\phi$ all higher order corrections. One can prove that this is indeed so provided
\begin{equation}
   \delta\phi = \kappa^2\lambda\int d^4x'd^4x''G(x-x')[G(x'-x'')]^3j(x')+O(j(x)^3)
\end{equation}
with the identification $\kappa=\mu$, the same of the exact solution, and $\bo G(x-x')+\lambda[G(x-x')]^3=\mu^{-1}\delta^4(x-x')$. This implies that the corresponding quantum field theory, in a very strong coupling limit, takes a Gaussian form and is trivial (triviality of the scalar field theory in the infrared limit). All we need now is to find the exact form of the propagator $G(x-x')$ and we have completely solved the classical theory for the scalar field in a strong coupling limit. In order to solve the equation
\begin{equation}
   \bo G(x-x')+\lambda[G(x-x')]^3=\mu^{-1}\delta^4(x-x')
\end{equation}
we can start from $d=1+0$ case $\partial_t^2G_0(t-t')+\lambda[G_0(t-t')]^3=\mu^2\delta(t-t')$. The Fourier transformed solution is straightforwardly obtained as
\begin{equation}
   G_0(\omega)=\sum_{n=0}^\infty(2n+1)\frac{\pi^2}{K^2(i)}\frac{(-1)^{n}e^{-(n+\frac{1}{2})\pi}}{1+e^{-(2n+1)\pi}}
   \frac{1}{\omega^2-m_n^2+i\epsilon}
\end{equation}
being $m_n=(2n+1)\frac{\pi}{2K(i)}\left(\frac{\lambda}{2}\right)^{\frac{1}{4}}\mu$ and $K(i)\approx 1.3111028777$ an elliptic integral. We are able to recover the full covariant propagator by boosting from the rest reference frame obtaining finally
\begin{equation}
   G(p)=\sum_{n=0}^\infty(2n+1)\frac{\pi^2}{K^2(i)}\frac{(-1)^{n}e^{-(n+\frac{1}{2})\pi}}{1+e^{-(2n+1)\pi}}
   \frac{1}{p^2-m_n^2+i\epsilon}.
\end{equation}
This shows that our solution given above indeed represents a strong coupling expansion being meaningful for $\lambda\rightarrow\infty$.

Similarly, for a classical Yang-Mills field one has to solve the set of nonlinear equations
\begin{eqnarray}
&&\partial^\mu\partial_\mu A^a_\nu-\left(1-\frac{1}{\alpha}\right)\partial_\nu(\partial^\mu A^a_\mu)+gf^{abc}A^{b\mu}(\partial_\mu A^c_\nu-\partial_\nu A^c_\mu)+gf^{abc}\partial^\mu(A^b_\mu A^c_\nu) \\ \nonumber
&&+g^2f^{abc}f^{cde}A^{b\mu}A^d_\mu A^e_\nu = -j^a_\nu.
\end{eqnarray}
For the homogeneous equation, we want to study it in the formal limit $g\rightarrow\infty$. We note that a class of exact solutions exists if we take the potential $A_\mu^a$ just depending on time, after a proper selection of the components (we dub this Smilga's choice \cite{Smilga:2001ck}). These solutions are the same of the scalar field when spatial coordinates are set to zero (rest frame). Differently from the scalar field, we cannot just boost away these solutions to get a general solution to Yang-Mills equations due to gauge symmetry. Anyhow, one can prove that the mapping persists but is just approximate in the limit of a very large coupling. This mapping would imply that we will have at our disposal a starting solution to build a quantum field theory for a strongly coupled Yang-Mills field. This solution displays a mass gap already at a classical level. Exactly as in the case of the scalar field we assume the following solution to our field equations
\begin{equation}
   A_\mu^a=\kappa\int d^4x'D_{\mu\nu}^{ab}(x-x')j^{b\nu}(x')+\delta A_\mu^a
\end{equation}
Also in this case, apart from a possible correction, this boils down to an expansion in powers of the currents as already guessed in the '80 \cite{Cahill:1985mh}. This implies that the corresponding quantum theory, in a very strong coupling limit, takes a Gaussian form and is trivial. The crucial point, as already pointed out in the eighties \cite{Cahill:1985mh,Goldman:1980ww}, is the exact determination of the gluon propagator in the low-energy limit. This is possible using the following mapping theorem:

{\bf MAPPING THEOREM:} {\sl An extremum of the action
\begin{equation}
\nonumber
    S = \int d^4x\left[\frac{1}{2}(\partial\phi)^2-\frac{\lambda}{4}\phi^4\right]
\end{equation}
is also an extremum of the SU(N) Yang-Mills Lagrangian when one properly chooses $A_\mu^a$ with some components being zero and all others being equal, and $\lambda=Ng^2$, being $g$ the coupling constant of the Yang-Mills field, when only time dependence is retained. In the most general case the following mapping holds
\begin{equation}
\nonumber
    A_\mu^a(x)=\eta_\mu^a\phi(x)+O(1/\sqrt{N}g),
\end{equation}
being $\eta_\mu^a$ some constants properly chosen, that becomes exact for the Lorenz gauge.}

This theorem was proved in \cite{Frasca:2007uz} and, after considering a criticism by Terry Tao, in \cite{Frasca:2009yp}. Tao agreed with this latest proof \cite{Tao}.

The mapping theorem permits us to write down immediately the propagator for the Yang-Mills equations in the Landau gauge for SU(N):
\begin{equation}
    D_{\mu\nu}^{ab}(p)=\delta_{ab}\left(\eta_{\mu\nu}-\frac{p_\mu p_\nu}{p^2}\right)\sum_{n=0}^\infty\frac{B_n}{p^2-m_n^2+i\epsilon}
    +O\left(\frac{1}{\sqrt{N}g}\right)
\end{equation}
being
\begin{equation}
    B_n=(2n+1)\frac{\pi^2}{K^2(i)}\frac{(-1)^{n+1}e^{-(n+\frac{1}{2})\pi}}{1+e^{-(2n+1)\pi}}
\end{equation}
and
\begin{equation}
    m_n=(2n+1)\frac{\pi}{2K(i)}\left(\frac{Ng^2}{2}\right)^{\frac{1}{4}}\Lambda.
\end{equation}
The constant $\Lambda$ must be the same constant that appears in the ultraviolet limit by dimensional transmutation, here arises as an integration constant \cite{Frasca:2010iv}. This is the propagator of a massive field theory and the mass poles arise dynamically from the non-linearities in the equations of motion. But, we are working classically yet and we cannot claim anything about the quantum theory at this stage. 

\section{Quantum field theory}

We can formulate a quantum field theory for the scalar field starting from the generating functional
\begin{equation}
     Z[j]=\int[d\phi]\exp\left[i\int d^4x\left(\frac{1}{2}(\partial\phi)^2-\frac{\lambda}{4}\phi^4+j\phi\right)\right].
\end{equation}
We can rescale the space-time variable as $x\rightarrow\sqrt{\lambda}x$ and rewrite the functional as
\begin{equation}
     Z[j]=\int[d\phi]\exp\left[i\frac{1}{\lambda}\int d^4x\left(\frac{1}{2}(\partial\phi)^2-\frac{1}{4}\phi^4+\frac{1}{\lambda}j\phi\right)\right].
\end{equation}
Then we can seek for a solution series as $\phi=\sum_{n=0}^\infty\lambda^{-n}\phi_n$ and rescale the current $j\rightarrow j/\lambda$ being this arbitrary. It is not difficult to see that the leading order correction can be computed solving the classical equation
\begin{equation}
     \bo\phi_0+\phi_0^3=j
\end{equation}
that we already know how to manage. This is completely consistent with our preceding formulation \cite{Frasca:2005sx} but now all is fully covariant. We are just using our ability to solve the classical theory. Using the approximation holding at strong coupling
\begin{equation}
     \phi_0=\mu\int d^4xG(x-x')j(x')+\ldots
\end{equation}
it is not difficult to write the generating functional at the leading order in a Gaussian form
\begin{equation}
     Z_0[j]=\exp\left[\frac{i}{2}\int d^4x'd^4x''j(x')G(x'-x'')j(x'')\right].
\end{equation}
This conclusion is really important: It says that the scalar field theory in d=3+1 is \underline{trivial} in the infrared limit. This functional describes a set of free particles with a mass spectrum
\begin{equation}
     m_n=(2n+1)\frac{\pi}{2K(i)}\left(\frac{\lambda}{2}\right)^\frac{1}{4}\mu
\end{equation}
that are the poles of the propagator, the one of the classical theory. We note that this propagator is describing free massive particles with a superimposed spectrum of a harmonic oscillator as they would have a structure.

For Yang-Mills theory, the generating functional can be written down with the following terms in the action
\begin{equation}
     S_{YM}=-\frac{1}{4}\int d^4x{\rm Tr}F^2-\frac{1}{2\alpha}\int d^4x(\partial\cdot A)^2,
\end{equation}
the ghost field
\begin{equation}
     S_g=-\int d^4x(\bar c^a\partial_\mu\partial^\mu c^a+g\bar c^a f^{abc}\partial_\mu A^{b\mu}c^c)
\end{equation}
and the corresponding current terms
\begin{equation}
     S_c=\int d^4x j^a_\mu(x)A^{a\mu}(x)+\int d^4x\left[\bar c^a(x)\varepsilon^a(x)+\bar\varepsilon^a(x) c^a(x)\right].
\end{equation}
We now use the mapping theorem fixing the form of the propagator in the infrared, \eg\ in the Landau gauge, as
\begin{equation}
     D_{\mu\nu}^{ab}(p)=\delta_{ab}\left(\eta_{\mu\nu}-\frac{p_\mu p_\nu}{p^2}\right)\sum_{n=0}^\infty\frac{B_n}{p^2-m_n^2+i\epsilon}
     +O\left(\frac{1}{\sqrt{N}g}\right)
\end{equation}
but this can be recomputed in any gauge by the classical equations with the mapping theorem. The next step is to use the approximation that holds in a strong coupling limit
\begin{equation}     
     A_\mu^a=\Lambda\int d^4x' D_{\mu\nu}^{ab}(x-x')j^{b\nu}(x')+O\left(\frac{1}{\sqrt{N}g}\right)+O(j^3)
\end{equation}
and we note that, in this approximation, the ghost field just decouples and becomes free and one finally has at the leading order
\begin{equation}
     Z_0[j]=\exp\left[\frac{i}{2}\int d^4x'd^4x''j^{a\mu}(x')D_{\mu\nu}^{ab}(x'-x'')j^{b\nu}(x'')\right].
\end{equation}
This functional describes free massive glueballs that are the proper states in the infrared limit. Yang-Mills theory is \underline{trivial} in the limit of the coupling going to infinity and we expect the running coupling to go to zero lowering energies. Now, we can take a look at the ghost part of the action. We just note that, for this particular form of the propagator, inserting our approximation into the action produces an action for a free ghost field. Indeed, we will have
\begin{equation}
      S_g=-\int d^4x\left[\bar c^a\partial_\mu\partial^\mu c^a+O\left(\frac{1}{\sqrt{N}g}\right)+O\left(j^3\right)\right].
\end{equation}
So, a ghost propagator can be written down as
\begin{equation}
      G_{ab}(p)=-\frac{\delta_{ab}}{p^2+i\epsilon}+O\left(\frac{1}{\sqrt{N}g}\right).
\end{equation}
Our conclusion is that, in a strong coupling expansion $1/\sqrt{N}g$, we get the so called \underline{decoupling solution}.

\section{QCD at infrared limit}

When use is made of the infrared fixed point result, QCD action can be written down quite easily. Indeed, we will have for the gluon field
\begin{equation}
      S_{gf}=\frac{1}{2}\int d^4x'd^4x''\left[j^{\mu a}(x')D_{\mu\nu}^{ab}(x'-x'')j^{\nu b}(x'')
      +O\left(\frac{1}{\sqrt{N}g}\right)+O\left(j^3\right)\right]
\end{equation}
and for the quark fields
\begin{eqnarray}
      S_q&=&\sum_q\int d^4x\bar q(x)\left[i{\slashed\partial}-m_q-g\gamma^\mu\frac{\lambda^a}{2}\int d^4x'D_{\mu\nu}^{ab}(x-x')j^{\nu b}(x')\right. \\ \nonumber
      &-&\left.g^2\gamma^\mu\frac{\lambda^a}{2}\int d^4x'D_{\mu\nu}^{ab}(x-x')\sum_{q'}\bar q'(x')\frac{\lambda^b}{2}\gamma^\nu q'(x')
      +O\left(\frac{1}{\sqrt{N}g}\right)+O\left(j^3\right)\right]q(x)
\end{eqnarray}
We recognize here an explicit Yukawa interaction and a Nambu-Jona-Lasinio non-local term. Already at this stage we are able to recognize that NJL is the proper low-energy limit for QCD at zero temperature. Now we operate the Smilga's choice $\eta_\mu^a\eta_\nu^b=\delta_{ab}(\eta_{\mu\nu}-p_\mu p_\nu/p^2)$ for the Landau gauge. We are left with the infrared limit QCD as one has using conservation of currents
\begin{equation}
      S_{gf}=\frac{1}{2}\int d^4x'd^4x''\left[j_\mu^a(x')G(x'-x'')j^{\mu a}(x'')
      +O\left(\frac{1}{\sqrt{N}g}\right)+O\left(j^3\right)\right]
\end{equation}
and for the quark fields
\begin{eqnarray}
      S_q&=&\sum_q\int d^4x\bar q(x)\left[i{\slashed\partial}-m_q-g\gamma^\mu\frac{\lambda^a}{2}\int d^4x'G(x-x')j^a_\mu(x')\right. \\ \nonumber
      &-&\left.g^2\gamma^\mu\frac{\lambda^a}{2}\int d^4x'G(x-x')\sum_{q'}\bar q'(x')\frac{\lambda^a}{2}\gamma_\mu q'(x')
      +O\left(\frac{1}{\sqrt{N}g}\right)+O\left(j^3\right)\right]q(x).
\end{eqnarray}
This action can be the starting point for our analysis at finite temperature. But before doing this, we want to give explicitly the contributions from gluon resonances. In order to do this, we introduce the bosonic currents $j^a_\mu(x)=\eta^a_\mu j(x)$ with the current $j(x)$ that of the gluonic excitations after mapping. Using the relation $\eta_\mu^a\eta^{\mu a}=3(N_c^2-1)$ we get in the end
\begin{equation}
      S_{gf}=\frac{3}{2}(N_c^2-1)\int d^4x'd^4x''\left[j(x')G(x'-x'')j(x'')
      +O\left(\frac{1}{\sqrt{N}g}\right)+O\left(j^3\right)\right]
\end{equation}
and for the quark fields
\begin{eqnarray}
      S_q&=&\sum_q\int d^4x\bar q(x)\left[i{\slashed\partial}-m_q-g\eta_\mu^a\gamma^\mu\frac{\lambda^a}{2}\int d^4x'G(x-x')j(x')\right. \\ \nonumber
      &-&\left.g^2\gamma^\mu\frac{\lambda^a}{2}\int d^4x'G(x-x')\sum_{q'}\bar q'(x')\frac{\lambda^a}{2}\gamma_\mu q'(x')
      +O\left(\frac{1}{\sqrt{N}g}\right)+O\left(j^3\right)\right]q(x).
\end{eqnarray}
Now, we recognize that the propagator is just a sum of Yukawa propagators weighted by exponential damping terms. So, we introduce the $\sigma$ field and truncate at the first excitation. This is a somewhat rough approximation but will be helpful in the following analysis. This means that we can write the bosonic currents contribution as coming from a boson field written down as $\sigma(x) = \sqrt{3(N_c^2-1)/B_0}\int d^4x'\Delta(x-x')j(x')$. So, the model we consider for our finite temperature analysis, directly derived from QCD, is \cite{Hell:2008cc}
\begin{equation}
      S_\sigma=\int d^4x\left[\frac{1}{2}(\partial\sigma)^2-\frac{1}{2}m_0^2\sigma^2\right]
\end{equation}
and for the quark fields
\begin{eqnarray}
      S_q&=&\sum_q\int d^4x\bar q(x)\left[i{\slashed\partial}-m_q-g\sqrt{\frac{B_0}{3(N_c^2-1)}}
      \eta_\mu^a\gamma^\mu\frac{\lambda^a}{2}\sigma(x)\right. \\ \nonumber
      &-&\left.g^2\gamma^\mu\frac{\lambda^a}{2}\int d^4x'G(x-x')\sum_{q'}\bar q'(x')\frac{\lambda^a}{2}\gamma_\mu q'(x')
      +O\left(\frac{1}{\sqrt{N}g}\right)+O\left(j^3\right)\right]q(x).
\end{eqnarray}
Now, we are able to recover recover the non-local model of Weise \etal \cite{Hell:2008cc} directly from QCD ($2{\cal G}(0)=G$ is the standard Nambu-Jona-Lasinio coupling) by setting
\begin{equation}
      {\cal G}(p)=-\frac{1}{2}g^2\sum_{n=0}^\infty\frac{B_n}{p^2-(2n+1)^2(\pi/2K(i))^2\sigma+i\epsilon}=\frac{G}{2}{\cal C}(p)
\end{equation}
with ${\cal C}(0)=1$ fixing in this way the value of $G$ using the gluon propagator.

We move to an Euclidean action and define the following fields
\begin{equation}     
   \phi_a(x)=(\sigma(x),{\bm\pi}(x)) 
\end{equation}
%being $\mu^{-2}(\omega)=\sum_{n=0}^\infty\frac{B_n}{\omega^2+m_n^2}$.
So, the bosonic action will be, after taking the expansion around the v.e.v. $\phi_a=(v,0)$,
\begin{equation}
      S_B=\int d^4x\left[\frac{1}{2}(\partial\delta\sigma)^2-\frac{1}{2}m_0^2(\delta\sigma)^2\right]+S_{MF}+S^{(2)}+\ldots      
\end{equation}
being
\begin{equation}
      S_{MF}/V_4=-2NN_f\int\frac{d^4p}{(2\pi^4}\ln\left[p^2+M^2(p)\right]+\frac{1}{2}\left(\frac{1}{G}+m_0^2\right)v^2.
\end{equation}
This holds together with the gap equations
\begin{equation}
      M(p)=m_q+{\cal C}(p)v
\end{equation}
and
\begin{equation}
      v=\frac{4NN_f}{m_0^2+1/G}\int\frac{d^4p}{(2\pi)^4}{\cal C}(p)\frac{M(p)}{p^2+M^2(p)}.
\end{equation}
The next-to-leading order term is given by (a correction to mass $m_0$ for the $\sigma$ field)
\begin{equation}      
   S^{(2)}=\frac{1}{2}\int\frac{d^4p}{(2\pi)^4}\left[F_+(p^2)\delta\sigma(p)\delta\sigma(-p)
   +F_-(p^2)\delta{\bm\pi}(p)\delta{\bm\pi}(-p)\right] 
\end{equation}
being
\begin{equation}
   F_{\pm}(p^2)=\frac{1}{G}-4NN_f\int\frac{d^4q}{(2\pi)^4}{\cal C}(q){\cal C}(q+p)\frac{q\cdot(q+p)\mp M(q)M(q+p)}
   {[q^2+M^2(q)][(q+p)^2+M^2(q+p)]}.
\end{equation}
For the chiral condensate one has
\begin{equation}
    \langle\bar\psi\psi\rangle=-4NN_f\int\frac{d^4p}{(2\pi)^4}\left[\frac{M(p)}{p^2+M^2(p)}-\frac{m_q}{p^2+m_q^2}\right].
\end{equation}
Till now there are two novelties really implied with respect to the work of Weise \etal \cite{Hell:2008cc}: The model is exactly obtained from QCD and the expression of the form factor ${\cal C}(p)$ is properly fixed through the exact gluon propagator at infrared. The next step is to consider the case of finite temperature. This can be easily accomplished with the exchange
\begin{equation}      
   \int\frac{d^4p}{(2\pi)^4}\rightarrow\beta^{-1}\sum_{k=-\infty}^\infty\int\frac{d^3p}{(2\pi)^3}
\end{equation}
being the sum over k that on Matsubara frequencies $\omega_k=2k\pi/\beta$ for bosons and $\omega_k=(2k+1)\pi/\beta$ for fermions. 
So, we can write down the gap equations at finite temperature as
\begin{equation}
    M(\omega_k,{\bm p})=m_q+{\cal C}(\omega_k,{\bm p})v
\end{equation}
\begin{equation}
    v=\frac{4NN_f}{m_0^2+1/G}\beta^{-1}\sum_{k=-\infty}^\infty\int\frac{d^3p}{(2\pi)^3}{\cal C}(\omega_k,{\bm p})\frac{M(\omega_k,{\bm p})}
    {\omega_k^2+{\bm p}^2+M^2(\omega_k,{\bm p})}
\end{equation}
while for the chiral condensate one has
\begin{equation}
    \langle\bar\psi\psi\rangle=-4NN_f\beta^{-1}\sum_{k=-\infty}^\infty\int\frac{d^3p}{(2\pi)^3}
    \left[\frac{M(\omega_k,{\bm p})}{\omega_k^2+{\bm p}^2+M^2(\omega_k,{\bm p})}-\frac{m_q}{\omega_k^2+{\bm p}^2+m_q^2}\right].
\end{equation}
Assuming the integral regularized by a cut-off $\Lambda$ and noting that ${\cal C}(p)$ is practically 1 in the low-energy range, we can prove the existence of a critical temperature where the chiral symmetry is restored. Setting $v=0$ into the gap equation we have to solve
\begin{equation}
    \frac{4NN_f}{m_0^2+1/G}\beta^{-1}\sum_{k=-\infty}^\infty\int\frac{d^3p}{(2\pi)^3}{\cal C}^2(\omega_k,{\bm p})\frac{1}
    {\omega_k^2+{\bm p}^2}=1.
\end{equation}
At small temperatures we are able to get the critical temperature
\begin{equation}
    T_c^2\approx\frac{3}{\pi^2}\left[\Lambda^2-\frac{\pi^2}{NN_f}\left(m_0^2+\frac{1}{G}\right)\right]
\end{equation}
This shows, starting directly from QCD, that a critical point does exist for this theory. We note that for $N_f=2$ and $T_c=170\ MeV$ gives $\Lambda=769\ MeV$, perfectly consistent with NJL model.
This expression is very similar to the one obtained in \cite{GomezDumm:2004sr}.

\begin{figure}[htb]
\begin{center}
\epsfig{file=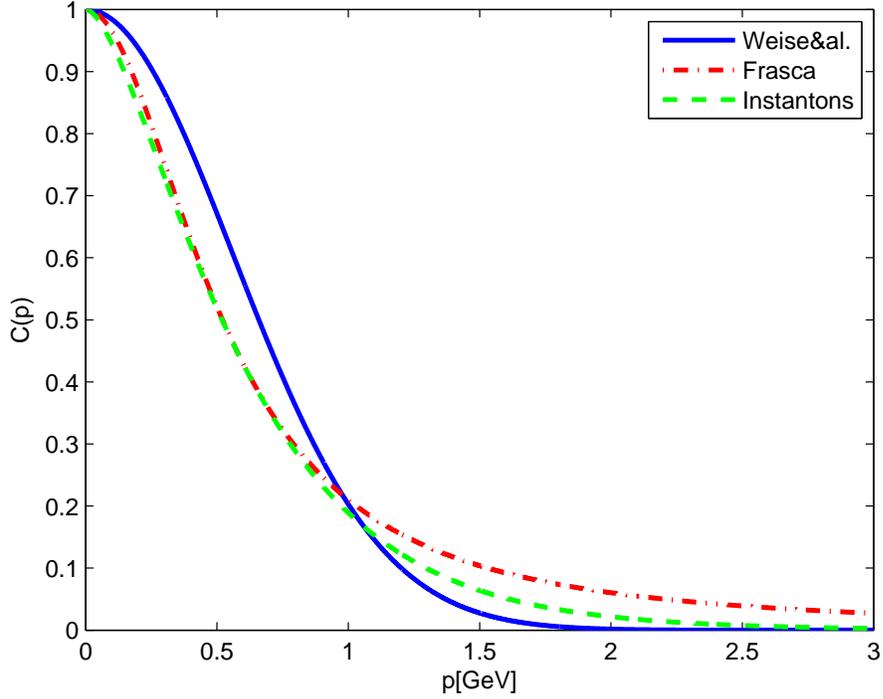,height=4in}
\caption{Comparison between the form factors with respect to the case of an instanton liquid.}
\label{fig:comp}
\end{center}
\end{figure}

For aims of completeness, we give in fig.\ref{fig:comp} a comparison of our gluon propagator (the form factor) with that used in Weise \etal \cite{Hell:2008cc} and the exact one for the instanton liquid as given in \cite{Schafer:1996wv} that inspired Weise \etal.
%\vspace{-0.6cm}
%\vspace{-0.8cm}
This shows that the approximation of a QCD vacuum as an instanton liquid is a very good one.

\bigskip
I am much grateful to Marco Ruggieri without whose help I would not have obtained these results.

%\def\Discussion{
%\setlength{\parskip}{0.3cm}\setlength{\parindent}{0.0cm}
%     \bigskip\bigskip      {\Large {\bf Discussion}} \bigskip}
%\def\speaker#1{{\bf #1:}\ }
%\def\endDiscussion{}

%\Discussion

%None.

%\endDiscussion
 
\end{document}